\documentclass[notitlepage,prd,preprintnumbers, longbibliography, showpacs]{revtex4-2}

\usepackage[colorlinks=true, allcolors=blue]{hyperref}
\usepackage{amsmath, mathtools}
\usepackage{multirow}
\usepackage{slashed}
\usepackage{braket}
\usepackage{subcaption}
\usepackage{booktabs}
\usepackage{graphicx}
\usepackage{cleveref}
\usepackage{float}
\usepackage{bm}
\graphicspath{{FIG/}}
\crefname{figure}{Figure}{Figures}
\crefname{table}{Table}{Tables}
\allowdisplaybreaks
\begin{document}

\title{Investigating the internal structure of $X(6900)$ in the $2J/\psi$ decay channel}
\author{Duo-Duo Lu}
\author{Shao-Zhou Jiang}
\email{jsz@gxu.edu.cn}
\affiliation{Key Laboratory for Relativistic Astrophysics, School of Physical Science and Technology, Guangxi University, Nanning 530004, People's Republic of China}

\begin{abstract}
Assuming $X(6900)$ is a tetraquark state, the decay width of $X(6900)\to 2J/\psi$ is calculated in a covariant quark model,
with the diquark-antidiquark $[cc][\bar{c}\bar{c}]$ picture.
Two possible structures, vector-vector and axial-vector--axial-vector coupling, are investigated.
The result indicates that the axial-vector--axial-vector coupling is consistent with the experiments.
Additionally, as another application of the covariant quark model, the decay width of $X(6900)\to 2\eta_c$ is predicted to be $66\sim88$ keV.
\end{abstract}

\maketitle

\section{INTRODUCTION}

Since the experimental discovery of the first exotic hadron state $X(3872)$ in 2003 \cite{PhysRevLett.91.262001},
a large number of new hadronic states have been observed \cite{chen2022updated}.
These new states usually do not conform to the simple picture of mesons and baryons in the conventional
quark model, and they are therefore collectively known as exotic hadrons.
These states may include tetraquark states, pentaquark states, hadronic molecule states, hybrid configurations, and so on.

In 2020, the LHCb analyzed $pp$ collision data at the center-of-mass energies
of $\sqrt{s}=7$, 8, and 13 TeV,
with an integrated luminosity of 9 $\mathrm{fb}^{-1}$, to investigate the invariant mass spectrum
of $J/\psi$-pairs \cite{lhcb_collaboration_observation_2020}. There are two notable features that are observed. A narrow resonance-like peak, labeled $X(6900)$,
and a broader enhancement near the $2J/\psi$ threshold.
Both structures exhibit a statistical significance exceeding $5\sigma$.
The $X(6900)$ signal may represent a tetraquark state composed of four charm quarks,
as suggested by some theoretical models \cite{chen2017hunting, zhang2022spectrum, PhysRevD.103.034001, chen2020strong, PhysRevD.106.094019}.
In 2023, the ATLAS Collaboration conducted a search for possible
$cc\bar{c}\bar{c}$ tetraquark states decaying into pairs of charmonium particles,
with the final four-muon state \cite{PhysRevLett.131.151902}. The two decay channels were investigated,
\begin{equation*}
J/\psi+J/\psi \to 4\mu, \quad J/\psi+\psi(2S) \to 4\mu.
\end{equation*}
In 2024, the CMS Collaboration confirmed the $X(6900)$ state
in the $2J/\psi$ channel \cite{hayrapetyan2024new}.
All of these experimental results are listed in Table~\ref{tab:1}.
Recently, CMS has detected the $J^{PC}$ quantum number of $X(6900)$. They exclude $J=0,1$ and give $J^{PC}=2^{++}$ with a high degree of certainty \cite{cms2025determination}.

\begin{table}[ht]
\centering
\caption{
The mass of $X(6900)$ and the $X(6900)\to 2J/\psi$ decay width
from the different experiments. They are in units of MeV.
}
\begin{tabular}{llccc}
\hline\hline
Collaboration & Model & Mass  & Decay width
\\
\hline
\multirow{2}{*}{LHCb~\cite{lhcb_collaboration_observation_2020}}
& No-Interference & $6905 \pm 11 \pm 7$ & $80 \pm 19 \pm 33$\\
& Interference    & $6886 \pm 11 \pm 11$ & $168 \pm 33 \pm 69$\\[6pt]
\multirow{2}{*}{ATLAS~\cite{PhysRevLett.131.151902}}
& Model A & $6860 \pm 30^{+10}_{-20}$ & $110 \pm 50^{+20}_{-10}$\\
& Model B & $6910 \pm 10 \pm 10$      & $150 \pm 30 \pm 10$\\[6pt]
\multirow{2}{*}{CMS~\cite{hayrapetyan2024new}}
& Interference    & $6847^{+44+48}_{-28-20}$ & $191^{+66+25}_{-49-17}$\\
& No-Interference & $6927 \pm 9 \pm 4$ & $122^{+24}_{-21} \pm 18$\\
\hline\hline
\end{tabular}
\label{tab:1}
\end{table}

There is another story about the fully bottom tetraquark $X_b$. In 2017,
CMS made the first observation of $\Upsilon(1S)$ pair production
in proton-proton collisions. The mass of $X_b$ is about 18.4 GeV \cite{khachatryan2017observation}.
However, these structures were not confirmed by the later experiments
\cite{lhcb2018search, blekman2020measurement}.

In theory, there are various methods to study the fully heavy tetraquarks.
Before the experimental discovery of $X(6900)$ in 2020, the masses of the fully
heavy tetraquarks $bb\bar{b}\bar{b}$ and $cc\bar{c}\bar{c}$ had already been predicted
as early as 2017, using QCD sum rules within the diquark–antidiquark picture \cite{chen2017hunting}.
They have predicted the mass of the $J^{PC}=2^{++}$ particle, but its mass is inconsistent with
the above experimental results. Ref.~\cite{zhang2022spectrum} used the
nonrelativistic constituent quark model to study the mass spectrum of the $S$-wave fully heavy tetraquark states. It considered that $X(6900)$ may be not a ground $cc\bar{c}\bar{c}$ tetraquark state, but an excited state.
Ref.~\cite{bai2019beauty} considered heavy quarks as a non-relativistic many-body system, and used the
diffusion Monte Carlo method to calculate the mass of the $J^{PC}=0^{++}$ $bb\bar{b}\bar{b}$ states.
Ref.~\cite{PhysRevD.103.034001} studied the mass spectrum of the fully heavy tetraquark in an extended
chromomagnetic model.
Lattice nonrelativistic QCD was used to search for the existence of a $bb\bar{b}\bar{b}$ tetraquark bound state with a mass below the lowest noninteracting bottomonium-pair threshold \cite{PhysRevD.97.054505}.
The decay branch ratio of fully heavy tetraquarks was studied through the Fierz rearrangement, together with the QCD sum
rule \cite{chen2020strong, PhysRevD.106.094019}.
Ref. \cite{PhysRevD.109.056016} studied the electromagnetic and
hadronic decay widths of the $S$-wave fully heavy tetraquark through the nonrelativistic QCD.
Ref. \cite{PhysRevD.109.014006} studied the
decays of the fully beauty tetraquark to $B$ meson pairs using QCD sum rules.
There is also a lot of other research about fully heavy tetraquark states
\cite{chapon2022prospects, alkofer2005nucleon,
feng2021exclusive, huang2021inclusive, maciula2021mechanism, gonccalves2021fully, PhysRevD.102.116014,
PhysRevD.104.114029, bedolla2020spectrum, PhysRevD.103.116027, mutuk2021nonrelativistic,
PhysRevD.104.036016,wang2024two,yang2025strong,wang2017analysis,wang2022analysis,yu2023s,zhu2021fully,Chen:2024orv,Dong:2022sef,Dong:2025coi}.
More relevant works can be found in Ref. \cite{chen2022updated,Wang:2025sic}.

For $X(6900)$, it is generally regarded as a compact tetraquark state. For example, Ref.~\cite{WU2025100184} studies heavy flavor tetraquark systems within the framework of the quark potential model, finds that $X(6900)$ state is a compact tetraquark state, but not a meson molecule. Ref.~\cite{PhysRevD.103.034024} adopts three different theoretical approaches, which all lead to the similar conclusion that two-meson components do not play dominant roles in the X(6900). They consider $X(6900)$ is a compact four-charm-quark state or another microscopic degree of freedom. Hence, we consider $X(6900)$ as a tetraquark state but not a molecular state in this paper.

This paper studies the fully heavy tetraquark candidate $X(6900)$ through
calculating $X(6900)\to 2J/\psi$ decay width in the covariant quark model \cite{goerke2016four, dubnicka2010quark, dubnicka2011one, faessler2009semileptonic,
PhysRevD.60.094002, PhysRevD.73.094013, PhysRevD.56.348, ivanov1996electromagnetic}, in order to constrain the possible internal structure of $X(6900)$.
Although there are many possible decay channels,
only the $2J/\psi$ channel is considered, because this channel has been observed by the experiments mentioned above.
Since the quantum number of $X(6900)$ has already been given experimentally, i.e. $J^{PC}=2^{++}$ \cite{cms2025determination},
only the spin-2 possibility is considered.
For the spin-2 particle, there exist the two most likely structures, vector-vector coupling (V-V coupling) and axial-vector--axial-vector coupling (A-A coupling). Assuming $X(6900)$ is a tetraquark state, both of them will be considered, through calculating the corresponding decay widths.
Through comparison with the experiments, the internal coupling structure of $X(6900)$ will be predicted.
The result tends to the A-A coupling. This result can serve as a reference for the other research on $X(6900)$. It also means that the covariant quark model works well for $X(6900)$, and the model can further extend to the other exotic hadron state.

The paper is organized as follows. Section \ref{sec:2} will develop the nonlocal four-quark interpolating current for $X(6900)$.
It leads to a nonlocal effective Lagrangian that describes the interactions
between $X(6900)$ and its constituent quarks. The coupling strength between
$X(6900)$ and its quarks is derived using the compositeness condition.
In Section \ref{sec:3}, the decay amplitude and decay width for the process $X(6900)\to 2J/\psi$ are calculated for different structures.
In Section \ref{sec:4}, the numerical results of the decay width $X\to 2J/\psi$ are given.
Section  \ref{sec:41} applies the covariant quark model to $X(6900)\to 2\eta_c$, in order to predict its decay width.
Section \ref{sec:5} concludes with a summary and gives some discussions.

\section{THEORETICAL FRAMEWORK}\label{sec:2}
This section first gives a short introduction to the covariant quark model, and then mainly discusses how to calculate the decay width of $X(6900)\to 2J/\psi$.
This model is widely used to calculate hadron decays and electromagnetic properties
\cite{goerke2016four, dubnicka2010quark, dubnicka2011one, faessler2009semileptonic,
PhysRevD.60.094002, PhysRevD.73.094013, PhysRevD.56.348, ivanov1996electromagnetic}.
The effective interaction Lagrangian is
\begin{equation}
\mathcal{L}_{\mathrm{int}}=\mathcal{L}_{X}+\mathcal{L}_{J/\psi},\label{eq:Lint}
\end{equation}
For simplicity, we use the subscript $X$ to denote $X(6900)$ for short hereafter.
$\mathcal{L}_{X}$ represents the effective interaction
for the tetraquark $X(6900)$ coupling to the four-quark current.
$\mathcal{L}_{J/\psi}$ represents the effective interaction
for the vector meson $J/\psi$ coupling to two-quark current $c\bar{c}$.

\begin{figure}[htbp]
\centering
\subfloat[]{%
    \includegraphics[width=0.32\textwidth]{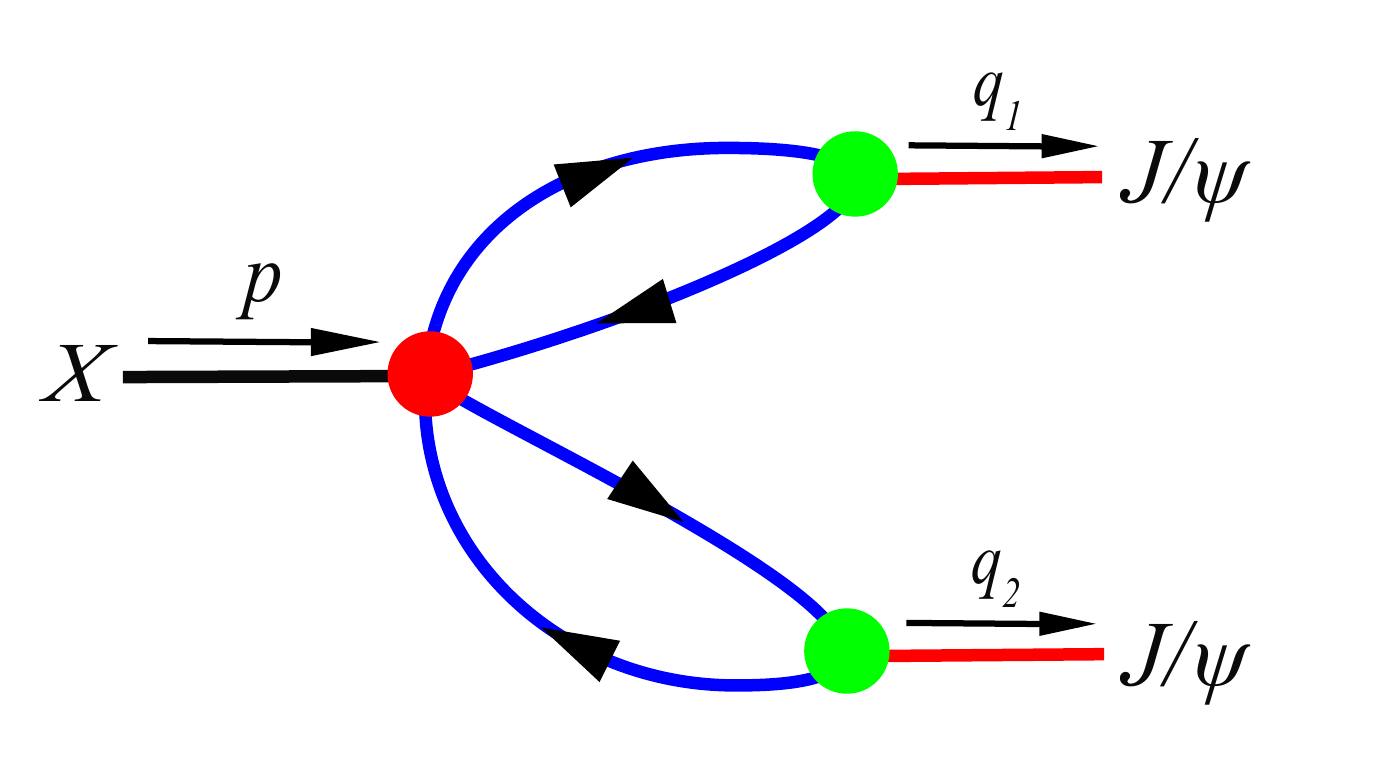}%
    \label{fig:X-decay1}%
}
\hfill
\subfloat[]{%
    \includegraphics[width=0.32\textwidth]{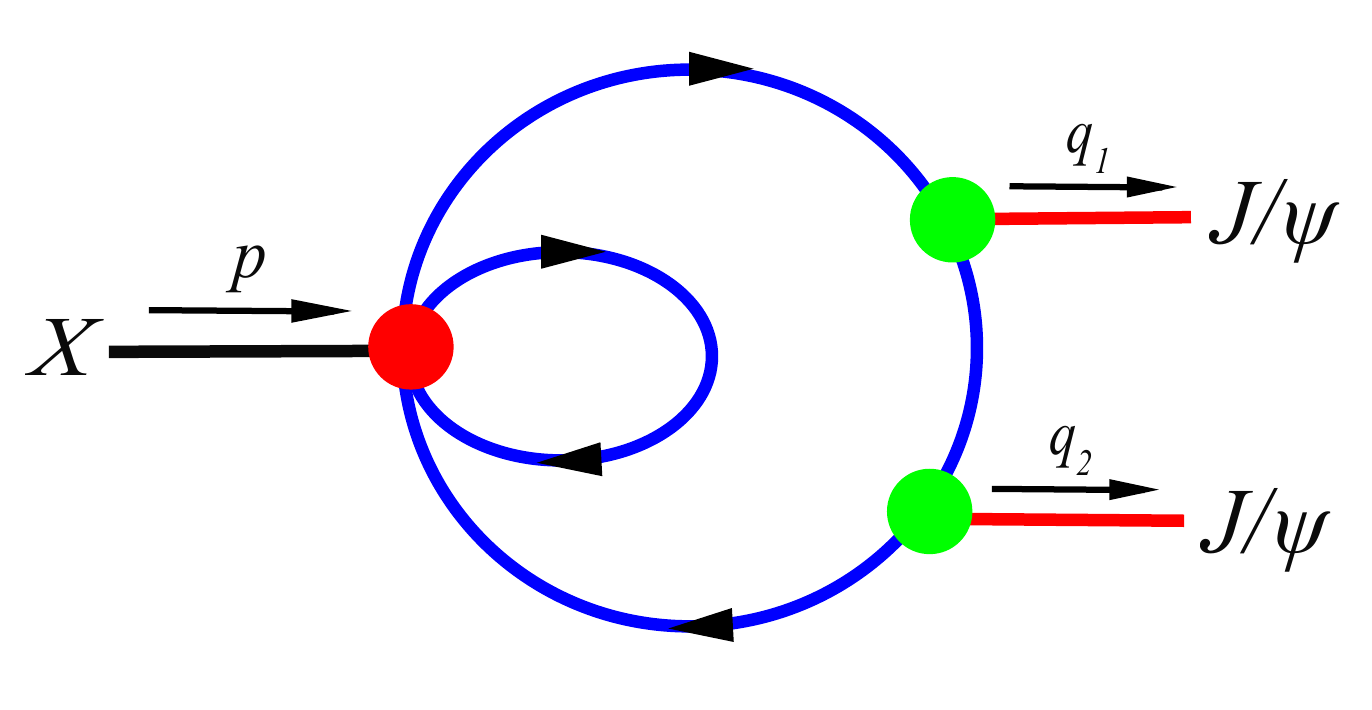}%
    \label{fig:X-decay2}%
}
\hfill
\subfloat[]{%
    \includegraphics[width=0.32\textwidth]{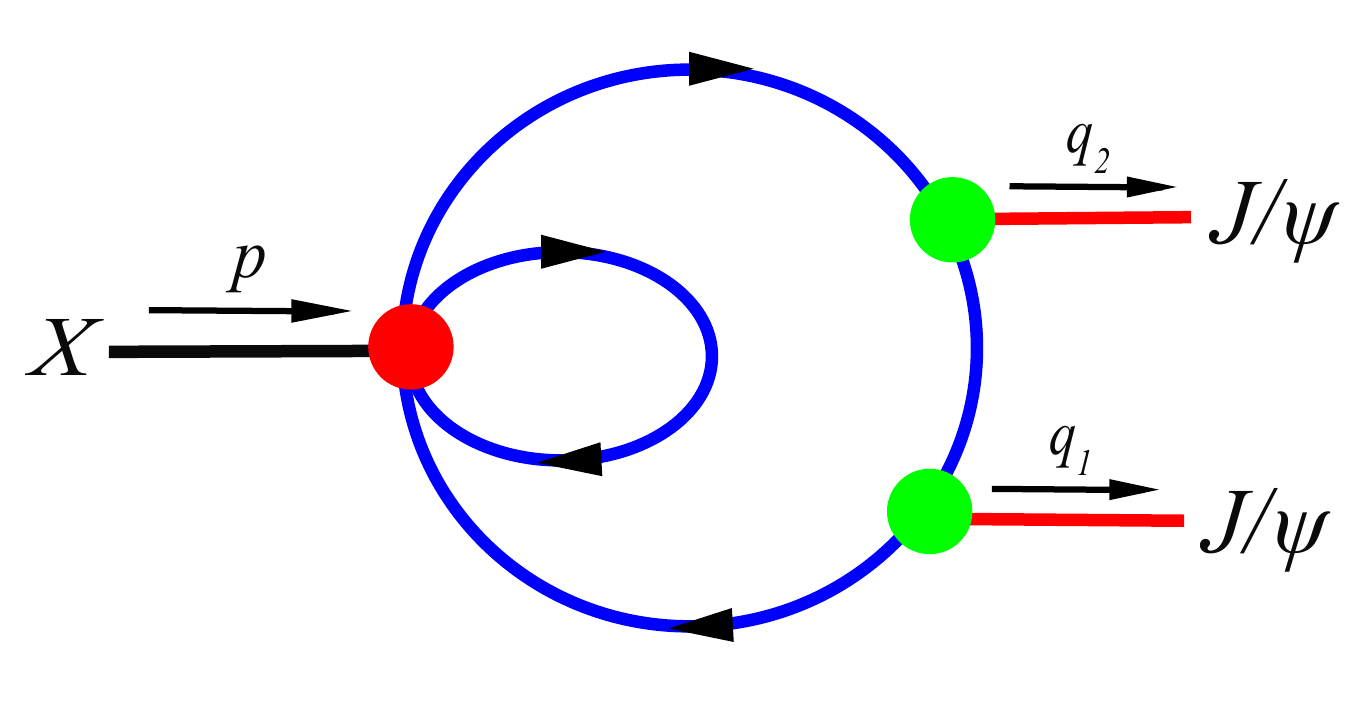}%
    \label{fig:X-decay3}%
}

\subfloat[]{%
  \includegraphics[width=0.32\textwidth]{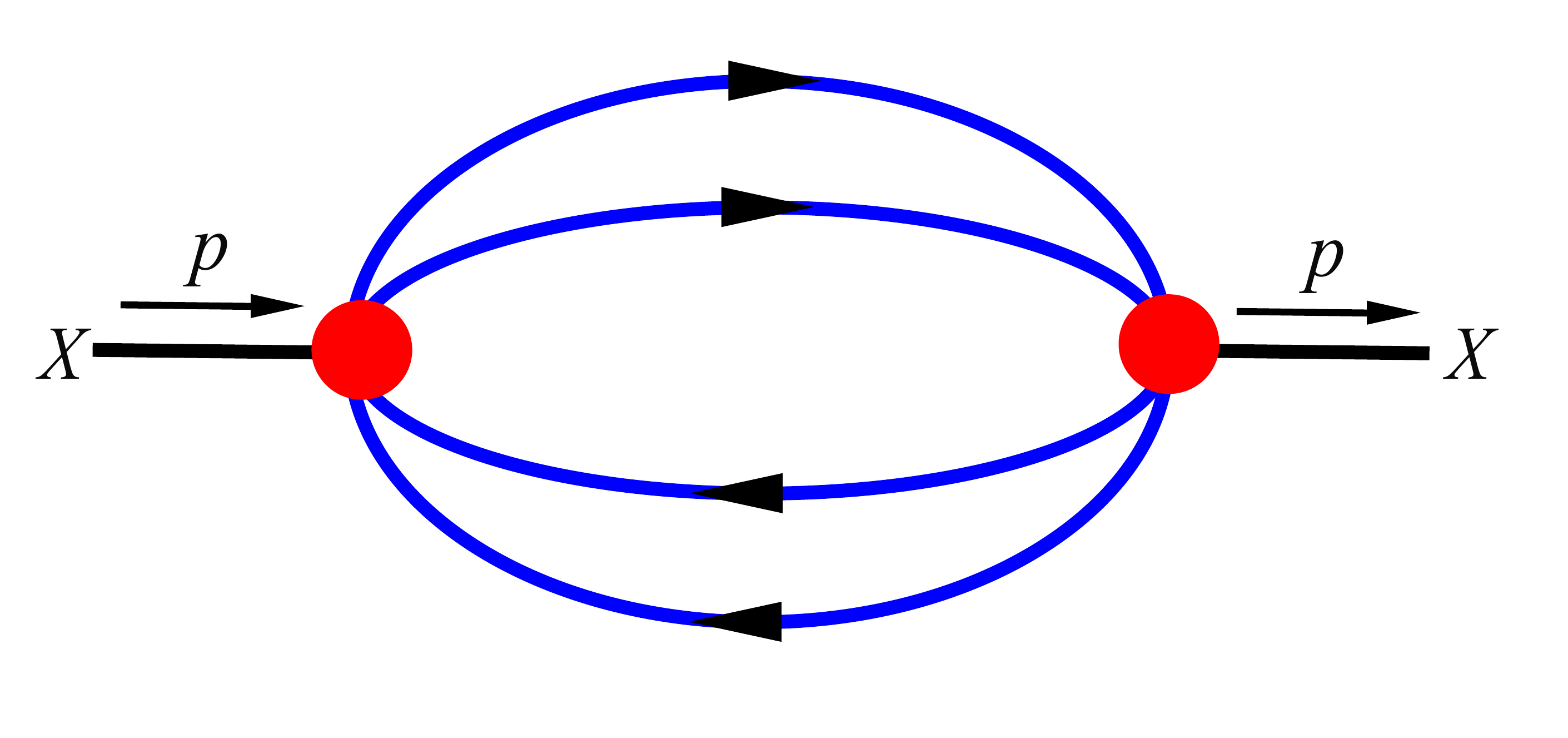}%
  \label{fig:X-3LOOP}%
}
\quad
\subfloat[]{%
  \includegraphics[width=0.32\textwidth]{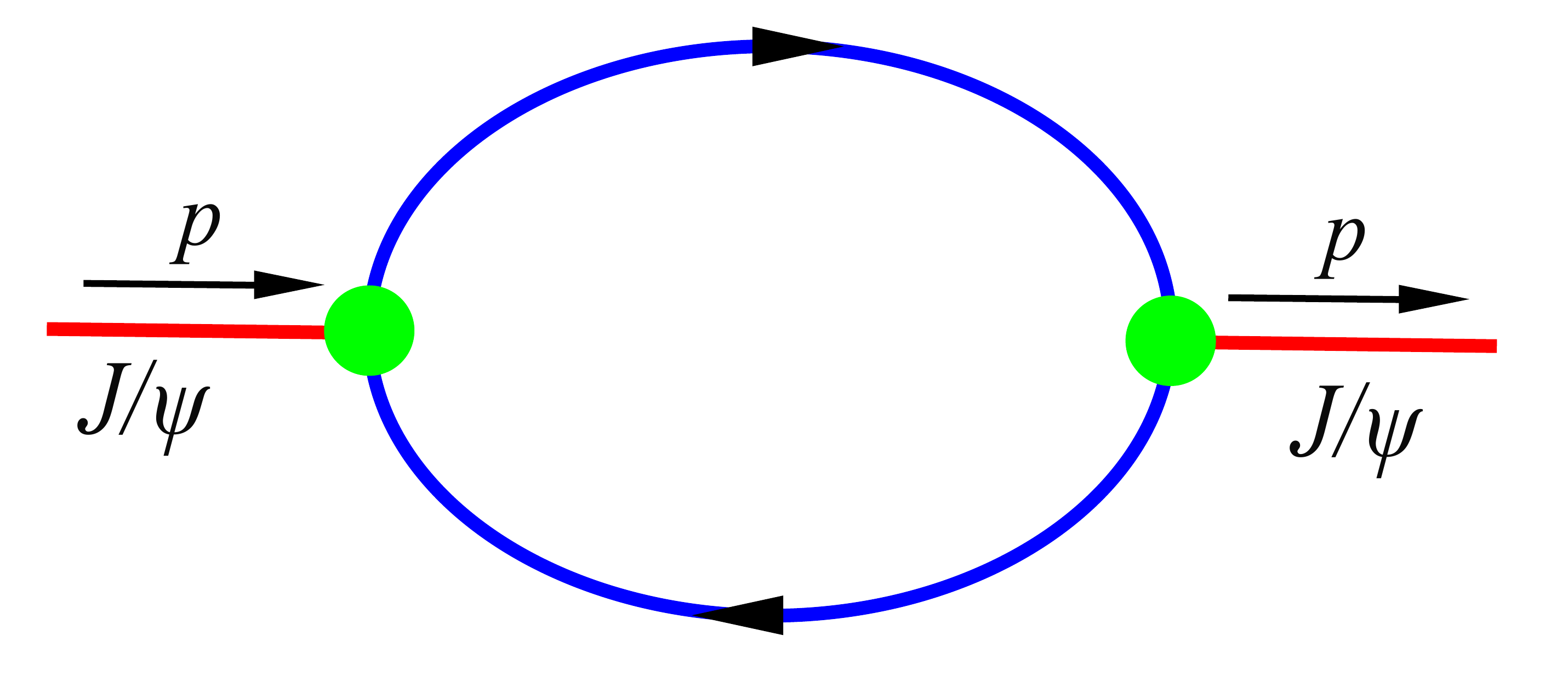}%
  \label{fig:J-psi}%
}
\caption{Feynman diagrams in this paper. (a), (b) and (c) are three $X \to 2J/\psi$ Feynman diagrams. (d) and (e) are the mass operators of $X(6900)$ ($\Pi_{X}^{\mu\nu,\alpha\beta}$) and $J/\psi$ ($\Pi_{J/\psi}^{\mu \nu}$), respectively. The red and green dots denote the interactions from $\mathcal{L}_{X}$ and $\mathcal{L}_{J/\psi}$, respectively.}
\label{fig:all}
\end{figure}

The lowest-order perturbation Feynman diagram of the decay $X(6900)\to 2J/\psi$ is showed in
Figures~\ref{fig:X-decay1}, \ref{fig:X-decay2} and~\ref{fig:X-decay3}.
$\mathcal{L}_{X}$ is related to the left vertex in these figures, while $\mathcal{L}_{J/\psi}$ is related to the right two vertices in
these figures. Section \ref{sec:LX} will discuss the interaction $\mathcal{L}_{X}$, and $\mathcal{L}_{J/\psi}$ will be discussed in Sec. \ref{eq:LJpsi}.

\subsection{$X(6900)$ coupling to the four-quark current}\label{sec:LX}
Because Ref. \cite{cms2025determination} suggested $J^{PC}=2^{++}$, we only consider $J^{PC}=2^{++}$ in this paper. For spin-2 $X(6900)$ particle, the effective Lagrangian $\mathcal{L}_{X}$ is
\begin{equation}
\mathcal{L}_{X}=g_{X}^{\prime}\,X_{\mu\nu}(x)\, J_{X}^{\mu\nu}(x),
\label{eq:lagrangian1}
\end{equation}
where $g_{X}^{\prime}$ is the coupling constant. The current $J^{\mu\nu}_{X}(x)$ is the interpolating four-quark current with $X(6900)$.
\begin{equation}
J_{X}^{\mu\nu}(x) = \int d^4x_1 \cdots \int d^4x_4 \, F_X(x, x_1, x_2, x_3,  x_4) j_{4c}^{\mu\nu}(x_1, x_2, x_3,  x_4).
\label{eq:current1}
\end{equation}
The specific form of $j_{4c}^{\mu\nu}$ depends on the internal structure of $X(6900)$.
It may possess several possible structures, including a diquark-antidiquark structure, a molecular configuration, and so on. According to the discussion in the Introduction, we suppose that it is a $[cc][\bar{c}\bar{c}]$ type tetraquark, which is also the mainstream
interpretation supported by most literature
\cite{chen2017hunting, zhang2022spectrum, PhysRevD.103.034001, chen2020strong, PhysRevD.106.094019}.
The four-quark current $j_{4c}^{\mu\nu}$ is defined as
\begin{equation}
   j_{4c}^{\mu\nu}(x) =
 [c_a^T(x)C \Gamma_1 c_b(x)]\,  [\bar{c}_a(x) \Gamma_2C \bar{c}_b^T(x)]
    + (\Gamma_1 \leftrightarrow \Gamma_2),
	\label{eq:j4c_local}
\end{equation}
where $C=i\gamma^2\gamma^0$ is the charge-conjugation matrix, $\Gamma_{1,2}$ are the elements of the Clifford algebra, $a$ and $b$ are the color indices,
and the superscript $T$ means the transpose in the spinor space.
In Ref.~\cite{chen2017hunting}, these currents are taken to be local.
In the present work, however, the currents are generalized to the nonlocal forms
\begin{equation}
   j_{4c}^{\mu\nu}(x_1, x_2, x_3,  x_4) =
 [c_a^T(x_1)C \Gamma_1 c_b(x_2)]\,  [\bar{c}_a(x_3) \Gamma_2C \bar{c}_b^T(x_4)]
    + (\Gamma_1 \leftrightarrow \Gamma_2).
	\label{eq:j4c_nonlocal}
\end{equation}
Different choices of $\Gamma_1,\Gamma_2$ lead to different currents. For the $J=2$ case, there exist two possible choices, i.e. the A-A and V-V couplings,
\begin{align}
\text{A-A coupling}: &\quad \Gamma_1=\gamma^{\mu},\;\Gamma_2=\gamma^{\nu},\\
\text{V-V coupling}: &\quad \Gamma_1=\gamma^{\mu}\gamma^{5},\;\Gamma_2=\gamma^{\nu}\gamma^{5}. \label{eq:VVc}
\end{align}

The vertex function $F_X$ in Eq. \eqref{eq:current1} is related to the scalar part of the Bethe-Salpeter amplitude and
characterizes the finite size of the tetraquark. It also relates to different space-time points. This is the reason why Eq. \eqref{eq:j4c_nonlocal} needs to be nonlocal.
The space-time translational invariance implies that the vertex function $F_X$ satisfies
\begin{equation}
F_X(x+a,x_1+a,x_2+a,x_3+a,x_4+a)=F_X(x,x_1,x_2,x_3,x_4),
\end{equation}
for any given four-vector $a$. The specific form
\begin{equation}
F_X(x, x_1, x_2, x_3, x_4)=\delta^{(4)}\bigg(
x - \sum_{i=1}^4 w_i x_i\bigg)
\Phi_{X}\bigg(\sum_{i<j}(x_i - x_j)^2\bigg)
\label{eq:FX}
\end{equation}
is often adopted \cite{faessler2009semileptonic}. The weights $w_i$ need to satisfy the normalization condition
\begin{equation}
\sum_{i=1}^{4} w_i=1.
\end{equation}
This paper takes them to be the mass ratio of quarks
\begin{equation}
w_i=m_i\bigg/{\sum_{j=1}^{4} m_j}.
\end{equation}
In the present case, all four (anti)charm quarks have the same mass, so $w_1=w_2=w_3=w_4=1/4$.

$\Phi_X$ in Eq. \eqref{eq:FX} is a nonlocal function involving the four quarks, but the arguments $x_i$ are not diagonal and they are hard to be integrated.
To simplify the following calculation, $\Phi_X$ needs to be diagonalized in advance.
The following linear transformation \cite{goerke2016four}
\begin{equation}
\begin{aligned}
x_1 &= x_c + \frac{2w_2 + w_3 + w_4}{2\sqrt{2}} \cdot y_1 - \frac{w_3 - w_4}{2\sqrt{2}} \cdot y_2 + \frac{w_3 + w_4}{2} \cdot y_3,  \\
x_2 &= x_c - \frac{2w_1 + w_3 + w_4}{2\sqrt{2}} \cdot y_1 - \frac{w_3 - w_4}{2\sqrt{2}} \cdot y_2 + \frac{w_3 + w_4}{2} \cdot y_3,  \\
x_3 &= x_c - \frac{w_1 - w_2}{2\sqrt{2}} \cdot y_1 + \frac{w_1 + w_2 + 2w_4}{2\sqrt{2}} \cdot y_2 - \frac{w_1 + w_2}{2} \cdot y_3,  \\
x_4 &= x_c - \frac{w_1 - w_2}{2\sqrt{2}} \cdot y_1 - \frac{w_1 + w_2 + 2w_3}{2\sqrt{2}}
\cdot y_2 - \frac{w_1 + w_2}{2} \cdot y_3,
\end{aligned}\label{eq:xy}
\end{equation}
leads to
\begin{equation}
\sum_{i<j}(x_i - x_j)^2=y_1^2+y_2^2+y_3^2.
\end{equation}
Hence, Eq. \eqref{eq:current1} is simplified to
\begin{equation}
\begin{aligned}
 J_{X}^{\mu\nu}(x) ={}&
\int d^4 y_1  \int d^4 y_2  \int d^4 y_3 \Phi_{X} (y_1^2 + y_2^2 + y_3^2)
\\
& \times  j_{4c}^{\mu\nu} \left[
x + \frac{y_1}{2 \sqrt{2}} + \frac{y_3}{4},  x - \frac{y_1}{2 \sqrt{2}}
+ \frac{y_3}{4},  x + \frac{y_2}{2 \sqrt{2}} - \frac{y_3}{4},  x - \frac{y_2}{2 \sqrt{2}} - \frac{y_3}{4}
\right].\label{eq:JX}
\end{aligned}
\end{equation}
There should be another $1/4$ factor from the Jacobian in Eq. \eqref{eq:JX}.
This factor has been absorbed in the new coupling constant $g_X=g_X^{\prime}/4$, in order to simplify the following calculation.

$\Phi_X$ in the momentum space is denoted to $\phi_X$, and
\begin{equation}
\Phi_X(y_1^2+y_2^2+y_3^2)=\left(\prod_{i=1}^{3} \int\dfrac{ d^4k_i}{(2\pi)^4}\right)
\phi_X(k_1^2+k_2^2+k_3^2) \,  e^{-i(k_1\cdot y_1+k_2\cdot y_2+k_3\cdot y_3)}.\label{eq:phik}
\end{equation}
In theory, the explicit form of $\phi_X$ can be derived by solving
the Bethe-Salpeter equation \cite{ivanov1999charm, alkofer2005nucleon}.
However, the strict form of $\phi_X$ is hard to calculate, but it needs to decrease rapidly enough in
the ultraviolet region in Euclidean space, in order to ensure that the associated Feynman diagrams remain
ultraviolet finite.
This paper chooses the Gaussian form,
\begin{equation}
\phi_X(k^2)=\exp(k^2/\Lambda_X^2).\label{eq:phiXk}
\end{equation}
This form also adopted in most literature \cite{goerke2016four, dubnicka2010quark, dubnicka2011one, faessler2009semileptonic,
PhysRevD.60.094002, PhysRevD.73.094013, PhysRevD.56.348, ivanov1996electromagnetic}.
Ref. \cite{anikin1995extended} studied different functional forms and found that the observable physical quantities are insensitive to the specific form of this vertex function.

The parameter $\Lambda_X$, which characterizes the size of $X(6900)$, is set around the mass of $X(6900)$~\cite{dubnicka2010quark,dubnicka2011one}. It is chosen within the range of $\Lambda_X=6.8 \sim 7.2 \, \mathrm{GeV}$.

\subsection{Coupling constant $g_X$}\label{sec:gX}
The coupling constant $g'_X$ in Eq. \eqref{eq:lagrangian1} or $g_X$ below Eq. \eqref{eq:JX} are unknown in the covariant quark model.
It can be derived by renormalizing the mass operator $\Pi_{X}$. $\Pi_{X}$ is defined by
\begin{equation}
\Pi_{X}^{\mu\nu,\alpha\beta}(p^2)=i \int d^4x \,  e^{ip \cdot x}
\langle 0 | TJ_X^{\mu\nu}(x)J_X^{\alpha\beta}(0)^\dagger | 0 \rangle .
\end{equation}
The corresponding Feynman diagram is shown in Fig. \ref{fig:X-3LOOP}.
For the A-A coupling,
\begin{equation}
\begin{aligned}
\Pi_{X}^{\mu\nu,\alpha\beta}(p^2)={}&\frac{12 i}{(2\pi)^{12}}  \,
\prod_{j=1}^4 \int d^4 p_j \,  \delta^{(4)}(p+p_1-p_2-p_3+p_4) \,
\phi_{X}^2
\left(
\frac{\tfrac{1}{2} (p_1 - p_2 + p_3 - p_4)^2 + (p_1 + p_2)^2 + (p_3 + p_4)^2 }{8 }
\right)
\\
& \times \Big\{
\mathrm{tr}\!\big[S(p_1)\,  \gamma^\mu\,  S(p_2) \, \gamma^\alpha\big]\,
\mathrm{tr}\!\big[S(p_3)\, \gamma^\beta  S(p_4) \gamma^\nu \big] +
\mu \leftrightarrow \nu
\Big\}
+\alpha \leftrightarrow \beta.
\end{aligned}
\label{eq:VPIX}
\end{equation}
For the V-V coupling,
\begin{equation}
\begin{aligned}
\Pi_{X}^{\mu\nu,\alpha\beta}(p^2)={}&\frac{24 i}{(2\pi)^{12}}  \,
\prod_{j=1}^4 \int d^4 p_j \,  \delta^{(4)}(p+p_1-p_2-p_3+p_4) \,
\phi_{X}^2
\left(
\frac{\tfrac{1}{2} (p_1 - p_2 + p_3 - p_4)^2 + (p_1 + p_2)^2 + (p_3 + p_4)^2 }{8 }
\right)
\\
& \times \Big\{
\mathrm{tr}\!\big[S(p_1) \gamma^\mu\gamma^5  S(p_2) \, \gamma^\alpha\gamma^5\big]
\mathrm{tr}\!\big[S(p_3)\gamma^\beta\gamma^5  S(p_4) \gamma^\nu\gamma^5 \big] +
\mu \leftrightarrow \nu
\Big\}
+\alpha \leftrightarrow \beta.
\end{aligned}
\end{equation}
where the Feynman propagator of the charm quark is
\begin{equation}
S(p) = \dfrac{i}{\slashed{p} - m_c+i\epsilon}.
\end{equation}
The coupling constant $g_X$ is obtained by the normalization
condition called the compositeness condition \cite{PhysRevD.81.034010}
\begin{equation}
Z_{X} = 1 - g_{X}^2 \,  \dfrac{d \Pi_{X}^{2}(s)}{ds}\bigg|_{s=m_{X}^2} = 0,
\end{equation}
where $\Pi_{X}^{2}$ is defined as
\begin{equation}
    \Pi_{X}^{\mu\nu,\alpha\beta}(p^2) =
    \left[
    \frac{1}{2}\left(G^{\mu\alpha}G^{\nu\beta}+G^{\mu\beta}G^{\nu\alpha}\right)
    -\frac{1}{3} G^{\mu\nu}G^{\alpha\beta}
    \right]
    \Pi_{X}^{2}(p^2) + \cdots,
\end{equation}
where ``..." is the other irrelevant items, and $G^{\mu\nu}$ is defined as
\begin{equation}
    G^{\mu\nu} = \frac{p^\mu p^\nu}{p^2} - g^{\mu\nu}.
\end{equation}

The analytical result of $g_X$ can be obtained directly. For the A-A coupling,
\begin{equation}
\begin{aligned}
	\dfrac{1}{g_X^2} =&\frac{3\Lambda_X^6}{128 \pi^6}
\left(\prod_{i=1}^4 \int_0^\infty d u_i \right)\,
\exp \left[
	\frac{F_3}{8F_2}M^2
	- F_0 m_1^2
	\right]
	\\
	&  \times
\dfrac{1}{F_2^7}	\Big[
	F_1^3F_3 M^4
+ 16 F_1^2 F_2(F_1 +F_3)M^2
   +32F_2^3F_4 (2F_1 +F_3)m_1^2\\
    &+8F_1F_2^2F_3F_4m_1^2 M^2
	+ 32 F_1F_2^2 (4F_1 +F_3)
	+16F_2^4F_3 m_1^4
	\Big].
\end{aligned}
\end{equation}
For the V-V coupling,
\begin{equation}
\begin{aligned}
	\dfrac{1}{g_X^2} =&\frac{3\Lambda_X^6}{64\pi^6}
\left(\prod_{i=1}^4 \int_0^\infty d u_i \right)\,
\exp \left[
	\frac{F_3}{8F_2}M^2
	- F_0 m_1^2
	\right]
	\\
	&  \times
\dfrac{1}{F_2^7}	\Big[
	F_1^3F_3 M^4
+ 16 F_1^2 F_2(F_1 +F_3)M^2
    -32F_2^3F_4 (2F_1 +F_3)m_1^2\\
    &-8F_1F_2^2F_3F_4m_1^2 M^2
	+ 32 F_1F_2^2 (4F_1 +F_3)
	+16F_2^4F_3 m_1^4
	\Big].
\end{aligned}
\end{equation}
The dimensionless mass parameter reads
\begin{equation}
M=\dfrac{m_{X}}{\Lambda_X},  \qquad m_1=\dfrac{m_c}{\Lambda_X},
\qquad m_2=\dfrac{m_{J/\psi}}{\Lambda_X}.
\end{equation}
The parameters $F_i$ are
\begin{equation}
\begin{aligned}
F_0 &=u_{1} + u_{2} + u_{3} + u_{4},
\\
F_1 &=(2u_1+1)(2u_2+1)(2u_3+1)(2u_4+1),
\\
F_2 &=4\left(u_{1}u_{2}u_{3} + u_{1}u_{2}u_{4} + u_{1}u_{4}u_{3} + u_{4}u_{2}u_{3}\right)
+ 4\left(u_{1}u_{2} + u_{1}u_{3} + u_{1}u_{4} + u_{2}u_{3} + u_{2}u_{4} + u_{3}u_{4}\right)
\\
&
\quad+ 3\left(u_{1} + u_{2} + u_{3} + u_{4}\right)
+ 2,
\\
F_3 &=32 u_{1}u_{2}u_{3}u_{4}
+ 12\left(u_{1}u_{2}u_{3} + u_{1}u_{2}u_{4} + u_{1}u_{4}u_{3} + u_{4}u_{2}u_{3}\right)
\\
&\quad
+ 4\left(u_{1}u_{2} + u_{1}u_{3} + u_{1}u_{4} + u_{2}u_{3} + u_{2}u_{4} + u_{3}u_{4}\right)
+ u_{1} + u_{2} + u_{3} + u_{4},
\\
F_4 &=2u_1u_2 + 2u_3u_4 + u_1 + u_2 + u_3 + u_4 + 1.
  	\end{aligned}
\end{equation}

\subsection{$J/\psi$ coupling to the two-quark current}\label{eq:LJpsi}

$\mathcal{L}_{J/\psi}$ in Eq. \eqref{eq:Lint}, is
\begin{equation}
\mathcal{L}_{J/\psi} = g_{J/\psi} \,  \psi_{\mu}(x) \, J_{J/\psi}^{\mu}(x),
\label{eq:lagrangian2}
\end{equation}
where $g_{J/\psi}$ is the coupling constant.
The current $J_{J/\psi}^{\mu}(x)$ is defined as \cite{dubnicka2011one}
\begin{equation}
J_{J/\psi}^{\mu}(x) = \int d^4x_1 \int d^4x_2 \,  \delta^{(4)}\left( x-\dfrac{x_1+x_2}{2}\right)
\Phi_{J/\psi}\left[ (x_1-x_2)^2 \right]\bar{c}_a(x_1)\,  \gamma^{\mu}\,  c_a(x_2).\label{eq:jjpsi}
\end{equation}
Similar to Eqs. \eqref{eq:phik} and \eqref{eq:phiXk}, $\phi_{J/\psi}$ denotes $\Phi_{J/\psi}$ in the momentum space,
\begin{equation}
\Phi_{J/\psi}(x^2)=\int \dfrac{ d^4k}{(2\pi)^4} \phi_{J/\psi}(k^2) \,  e^{-ik\cdot x}.
\end{equation}
$\phi_{J/\psi}$ is also chosen as a Gaussian form,
\begin{equation}
\phi_{J/\psi}(k^2)=\exp(k^2/\Lambda_{J/\psi}^2). \label{eq:phiJpsi}
\end{equation}
This paper chooses the fitted value $\Lambda_{J/\psi}=3.3$ GeV \cite{dubnicka2010quark,PhysRevD.81.034010}.

$g_{J/\psi}$ is also unknown. It can be obtained by a similar method in Sec. \ref{sec:gX}. The $J/\psi$ mass operator $\Pi_{J/\psi}^{\mu \nu}$ is
\begin{equation}
\Pi_{J/\psi}^{\mu \nu}(p^2)=i \int d^4x \,  e^{ip \cdot x}
\langle 0 | TJ_{J/\psi}^{\mu}(x)J_{J/\psi}^{\nu}(0)^{\dagger} | 0 \rangle,
\end{equation}
which is related to Fig. \ref{fig:J-psi}. It can be simplified to
\begin{equation}
\Pi_{J/\psi}^{\mu\nu}(p^2)=-\frac{3 i }{16 \pi^4}\ 	\int d^4 k \
\mathrm{tr}\!\left[
\gamma^{\nu} \,  S(k-p) \,  \gamma^{\mu} \,  S(k)  \right]
\ \phi^2_{J/\psi}\left[(p/2 - k)^2\right].
\end{equation}
The coupling constant $g_{J/\psi}$ is determined by the compositeness condition
\begin{equation}
1 - g_{J/\psi}^2 \,  \dfrac{d \Pi^1_{J/\psi}(s)}{ds}\bigg|_{s=m_{J/\psi}^2} = 0. \label{eq:rgJpsi}
\end{equation}
$\Pi^1_{J/\psi}(p^2)$ is defined as
\begin{equation}
\Pi_{J/\psi}^{\mu\nu}(p^2)=\bigg(\frac{p^\mu p^\nu}{p^2} - g^{\mu\nu}\bigg)\Pi^1_{J/\psi}(p^2)+\dfrac{p^{\mu}p^{\nu}}{p^2}\Pi^0_{J/\psi}(p^2).
\end{equation}
The analytical results of $g_{J/\psi}$ is
\begin{equation}
\begin{aligned}
	\dfrac{1}{g_{J/\psi}^2}=&
	\frac{3}{8 \pi^{2}}
\int_{0}^{\infty} d u_{1} \int_{0}^{\infty} d u_{2}
\ \frac{1}{ D_1^{5}}\
\left(
D_1^{2} D_2 \,  m_1^{\prime 2} + D_1 \left( D_2 + 2 D_3 \right) + D_2 D_3 \,  m_2^{\prime 2}
\right)
\exp\Big[\frac{D_2 }{D_1} \frac{m_2^{\prime 2}}{2} - m_1^{\prime 2} (u_1 + u_2)\Big],
\end{aligned}
\end{equation}
where the dimensionless mass parameters are
\begin{equation}
m_1^{\prime }=\dfrac{m_{c}}{\Lambda_{J/\psi}},  \quad
m_2^{\prime }=\dfrac{m_{J/\psi}}{\Lambda_{J/\psi}},
\end{equation}
and the $D_i$ parameters are
\begin{equation}
\begin{aligned}
D_1 &=u_{1} + u_{2} + 2,
\\
D_2 &=
2u_{1}u_{2}+ u_{1}+u_{2},
\\
D_3 &=(u_1+1)(u_2+1).
\end{aligned}
\end{equation}

\section{THE DECAY WIDTH OF $X(6900) \to 2J/\psi $}
\label{sec:3}

The scattering amplitude of the decay
$X(p) \to J/\psi(q_1) + J/\psi(q_2)$
can be calculated from Figures~\ref{fig:X-decay1}--\ref{fig:X-decay3}.
The amplitudes corresponding to these Feynman diagrams are denoted as
$\mathcal{M}_a$, $\mathcal{M}_b$ and $\mathcal{M}_c$, respectively.
All the scalar integrals $R_i,N_i$ defined below can be found in Appendix \ref{subsec:SI}.

For the A-A coupling,
\begin{equation}
\begin{aligned}
\mathcal{M} &= \mathcal{M}_a+\mathcal{M}_b+\mathcal{M}_c
\\
&=
\varepsilon_{\mu}^{J/\psi}(q_1)^{*}
\varepsilon_{\nu}^{J/\psi}(q_2)^{*}
\varepsilon_{\alpha\beta}^X(p)
\Big[
 -\Big(m_XN_1+\dfrac{N_2}{m_X} (q_1 \cdot q_2)\Big) g^{\alpha\mu} g^{\beta\nu}
+\dfrac{N_2}{m_X} q_1^{\alpha}\Big( q_2^{\mu} g^{\beta\nu}
- q_1^{\nu} g^{\beta\mu}
- q_2^{\beta} g^{\mu\nu}
\Big)
\Big],
\end{aligned}\label{eq:Mav-av}
\end{equation}
where $\varepsilon_{\mu}^{J/\psi}$ and $\varepsilon_{\alpha\beta}^X$ represent the polarization of $J/\psi$ and $X(6900)$ fields, respectively.
\begin{equation}
\begin{aligned}
\mathcal{M}_a =&\;
\frac{3}{32 \pi^{8}} \,  g_X g_{J/\psi}^{2} \,
\varepsilon_{\mu}^{J/\psi}(q_1)^{*}
\varepsilon_{\nu}^{J/\psi}(q_2)^{*}
\varepsilon_{\alpha\beta}^X(p)
\int d^4 p_1 \,  d^4 p_2
\\[1mm]
&\quad \times
\phi_{J/\psi}\!\left[ \Big(\dfrac{q_{1}}{2}- p_{1}\Big)^{2}\right]
\phi_{J/\psi}\!\left[ \Big(\dfrac{q_{2}}{2}- p_{2}\Big)^{2}\right]
\phi_X\!\left(
\dfrac{8 p_1^2 - 8 p_1 \cdot q_1 + 8 p_2^2 - 8 p_2 \cdot q_2 + 3 q_1^2 - 2 q_1 \cdot q_2 + 3 q_2^2}{16}
\right)
\\
&\quad \times
\mathrm{tr}\Big[S(p_1)  \gamma^\alpha S(p_2 - q_2) \gamma^\nu S(p_2) \gamma^{\beta} S(p_1 - q_1) \gamma^\mu\Big],
\end{aligned}
\end{equation}
\begin{equation}
\mathcal{M}_b =\mathcal{M}_c=0.
\end{equation}

For the V-V coupling,
\begin{equation}
\begin{aligned}
\mathcal{M} &= \mathcal{M}_a+\mathcal{M}_b+\mathcal{M}_c
\\
&=
\varepsilon_{\mu}^{J/\psi}(q_1)^{*}
\varepsilon_{\nu}^{J/\psi}(q_2)^{*}
\varepsilon_{\alpha\beta}^X(p)
\Big[
 -\Big(-m_XR_1+\dfrac{R_2}{m_X} (q_1 \cdot q_2)\Big) g^{\alpha\mu} g^{\beta\nu}
+\dfrac{R_2}{m_X} q_1^{\alpha}\Big( q_2^{\mu} g^{\beta\nu}
- q_1^{\nu} g^{\beta\mu}
- q_2^{\beta} g^{\mu\nu}
\Big)
\Big],
\end{aligned}
\end{equation}
\begin{equation}
\begin{aligned}
\mathcal{M}_a =&\;
-\frac{3}{16 \pi^{8}} \,  g_X g_{J/\psi}^{2} \,
\varepsilon_{\mu}^{J/\psi}(q_1)^{*}
\varepsilon_{\nu}^{J/\psi}(q_2)^{*}
\varepsilon_{\alpha\beta}^X(p)
\int d^4 p_1 \,  d^4 p_2
\\[1mm]
&\quad \times
\phi_{J/\psi}\!\left[ \Big(\dfrac{q_{1}}{2}- p_{1}\Big)^{2}\right]
\phi_{J/\psi}\!\left[ \Big(\dfrac{q_{2}}{2}- p_{2}\Big)^{2}\right]
\phi_X\!\left(
\dfrac{8 p_1^2 - 8 p_1 \cdot q_1 + 8 p_2^2 - 8 p_2 \cdot q_2 + 3 q_1^2 - 2 q_1 \cdot q_2 + 3 q_2^2}{16}
\right)
\\
&\quad \times
\mathrm{tr}\Big[S(p_1)  \gamma^\alpha\gamma^5 S(p_2 - q_2) \gamma^\nu S(p_2) \gamma^{\beta}\gamma^5 S(p_1 - q_1) \gamma^\mu\Big],
\end{aligned}
\end{equation}
\begin{equation}
\mathcal{M}_b =\mathcal{M}_c=0.\label{eq:Mbcv-v}
\end{equation}

The two-body decay width in the center-of-mass system of $X(6900) \to 2J/\psi$ is
\begin{equation}
\Gamma =\dfrac{1}{2} \frac{|\bm{p}_{J/\psi}|}{8\pi m_X^2} |\overline{\mathcal{M}}|^2,
\end{equation}
where the factor $1/2$ comes from the symmetry of the identical particles, and the momentum of the final particle
\begin{equation}
|\bm{p}_{J/\psi}|=\sqrt{m_X^2/4-m_{J/\psi}^2}.
\end{equation}

The decay width of the A-A coupling is
\begin{equation}
\begin{aligned}
\Gamma={}&
\frac{(m_X^2/4 - m_{J/\psi}^2)^{1/2}}{1920\pi }
\Bigg[
N_1^{2}\left(\dfrac{m_X^{4}}{m_{J/\psi}^{4}}+\dfrac{12 m_X^{2}}{m_{J/\psi}^{2}}+56\right)
\\
&
+80N_1 N_2 \left(1-\dfrac{m_{J/\psi}^{2}}{ m_X^{2}}\right)
+12N_2^{2}\left(1-3\dfrac{m_{J/\psi}^{2}}{ m_X^{2}}+6\dfrac{m_{J/\psi}^{4}}{ m_X^{4}}\right)
\Bigg].\label{eq:Gammaav}
\end{aligned}
\end{equation}
For the V-V coupling,
\begin{equation}
\begin{aligned}
\Gamma={}&
\frac{(m_X^2/4 - m_{J/\psi}^2)^{1/2}}{1920\pi }
\Bigg[
R_1^{2}\left(\dfrac{m_X^{4}}{m_{J/\psi}^{4}}+\dfrac{12 m_X^{2}}{m_{J/\psi}^{2}}+56\right)
\\
&
-80R_1 R_2 \left(1-\dfrac{m_{J/\psi}^{2}}{ m_X^{2}}\right)
+12R_2^{2}\left(1-3\dfrac{m_{J/\psi}^{2}}{ m_X^{2}}+6\dfrac{m_{J/\psi}^{4}}{ m_X^{4}}\right)
\Bigg].
\end{aligned}\label{eq:Gammav}
\end{equation}

\section{NUMERICAL RESULTS AND DISCUSSION}
\label{sec:4}

The input masses are $m_{X}=6.905$ GeV, $m_{J/\psi}=3.0969$ GeV \cite{ParticleDataGroup:2024cfk} and $m_c=1.67$ GeV.
Because $X(6900)$ may be an excited state~\cite{zhang2022spectrum}, $m_c$ here is the constituent quark mass but not the current quark mass~\cite{goerke2016four}.

\begin{figure}[htbp]
\centering
\includegraphics[width=0.4\textwidth]{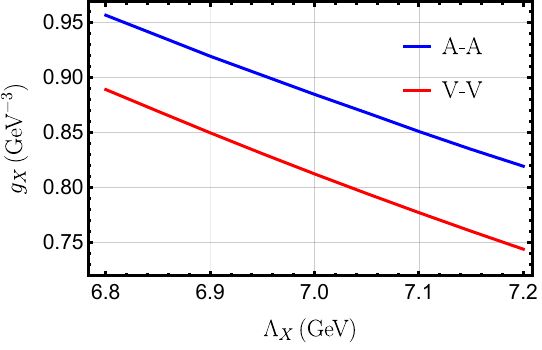}
\caption{Numerical results of $g_X$}
\label{fig:gx}
\end{figure}

The numerical result of $g_{J/\psi}$ is shown in \cref{tab:gj}. It indicates that $g_{J/\psi}$ decreases as $\Lambda_{J/\psi}$ increases, but the decrease of $g_{J/\psi}$ is slower than the increase of $\Lambda_{J/\psi}$. The dependency relationship between $\Lambda_X$ and $g_X$ is shown in \cref{fig:gx}. $g_X$ decreases as $\Lambda_{X}$ increases. The A-A coupling is a little larger than the V-V, but the difference is small. From the above results, it can be seen that the $g_{J/\psi}$ and $g_X$ coupling constants are weakly dependent on the input parameters $\Lambda_{J/\psi}$ and $\Lambda_{X}$, i.e. the model is not very sensitive to these inputs. This is what a good model requires.

\begin{table}[H]
\caption{Coupling $g_{J/\psi}$ for different $\Lambda_{J/\psi}$}
\centering
\setlength{\tabcolsep}{9pt}  
\begin{tabular}{l *{3}{c} }
    \hline\hline
    $\Lambda_{J/\psi}$ (GeV)  & 3.2 & 3.3 & 3.4
	\\
    \hline
    $g_{J/\psi}$     & 3.29   &   3.26   &   3.22
	\\
    \hline\hline
\end{tabular}
\label{tab:gj}
\end{table}

\begin{figure}[htbp]
\begin{minipage}[t]{0.9\textwidth}
    \begin{subfigure}{0.45\textwidth}
      \centering
      \includegraphics[width=\textwidth]{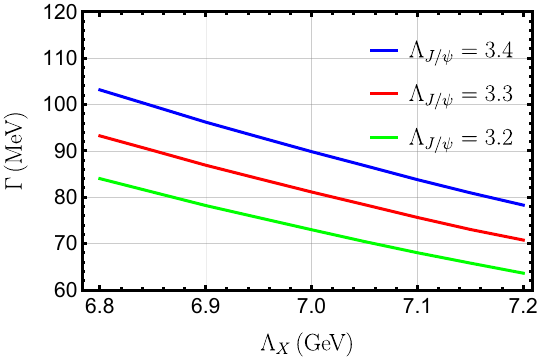}
      \caption{A-A coupling}
      \label{fig:2++}
  \end{subfigure}
  \hfill
    \begin{subfigure}{0.45\textwidth}
      \centering
      \includegraphics[width=\textwidth]{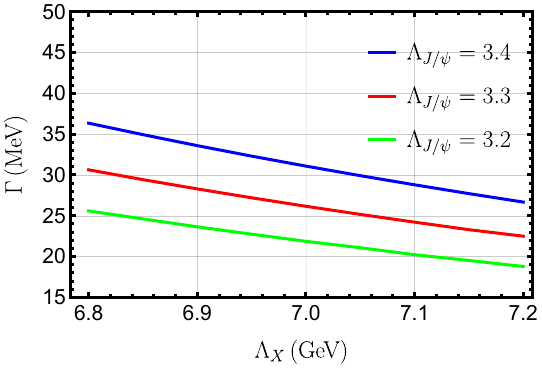}
      \caption{V-V coupling}
      \label{fig:2-+}
  \end{subfigure}
\caption{Decay width of $X(6900)\to 2J/\psi$.}
  \label{fig:width}
\end{minipage}
\end{figure}

The decay width of $X(6900)\to 2J/\psi$ is shown in \cref{fig:width}, with both A-A and V-V couplings.
To examine the sensitivity of the decay width to the parameter $\Lambda_{J/\psi}$,
we introduce a deviation of $\pm 0.1 \text{ GeV}$ around the fitted value $3.3 \text{ GeV}$.
The result of the A-A coupling is obviously larger than that from the V-V coupling and is consistent with the experiments, see Table~\ref{tab:1}. \cref{fig:width} also shows that $\Gamma_X$ is weakly dependent on both $\Lambda_X$ and $\Lambda_{J/\psi}$, for both couplings. Unless $\Lambda_X$ is very small and/or $\Lambda_{J/\psi}$ is very large, the V-V coupling cannot be consistent with the experiments. Table \ref{tab:gj} and \cref{fig:gx} have showed that the coupling constants change very little in the different inputs. Hence, comparing with the coupling constants, the main contribution comes from the way of how the four charm quarks couple. $X(6900)$ is more possible to have an internal structure with A-A coupling. Table~\ref{tab:com} presents a comparison between our result and those of others. Our result and the others are in the same order of magnitude. Compared to Table \ref{tab:1}, our result is better than Ref. \cite{agaev2024fully}, but gives a smaller value than Ref. \cite{PhysRevD.109.056016}. This may be because $\Lambda_X$ is chosen too small. It needs more experiments to fit the parameters.

\begin{table}[H]
\caption{A comparison between our results and those of others}
\label{tab:com}
\centering
\setlength{\tabcolsep}{9pt}  
\begin{tabular}{cccc}
    \hline\hline
    Physical quantity  &   Ours  &   Others
	\\
     \hline
    $\Gamma(X\to2J/\psi)$(MeV)  &   $71\sim93$  &   120 \cite{PhysRevD.109.056016}  &   56 \cite{agaev2024fully}
	\\
    $\Gamma(X\to2\eta_c)/\Gamma(X\to2J/\psi)(10^{-3})$    &   0.93     &  36 \cite{chen2020strong}  &  429 \cite{agaev2024fully}
	\\
    \hline\hline
\end{tabular}
\end{table}

Eqs. \eqref{eq:Mav-av}--\eqref{eq:Mbcv-v} tell us that only \cref{fig:X-decay1} has a contribution to the decay width. \cref{fig:X-decay2,fig:X-decay3} give zero contribution. This is because the quark loops from different quark pairs cancel,
such as $c_b(x_2), \bar{c}_a(x_3)$ and $c_a(x_1), \bar{c}_b(x_4)$ in Eq. \eqref{eq:j4c_nonlocal}. The distinction of the two decay widths mainly comes from the difference in the first sign in the second line in both Eqs. \eqref{eq:Gammaav} and \eqref{eq:Gammav}. 

\section{$X(6900)\to 2\eta_c$}\label{sec:41}
As another application of the covariant quark model, this section briefly discusses the decay width $X(6900)\to 2\eta_c$. At present, there does not exist any experiment about this process, but the covariant quark model can also predict its decay width. The Feynman diagrams are similar to Fig. \ref{fig:all}, except changing $J/\psi$ to $\eta_c$. The calculation in Sec. \ref{sec:2} can almost apply to $X(6900)\to 2\eta_c$.

\subsection{$\eta_c$ coupling to the two-quark current}\label{eq:Letac}

The effective interaction for the pseudoscalar meson $\eta_c$ coupling to the two-quark current $\bar{c}c$ is
\begin{equation}
\mathcal{L}_{\eta_c} = g_{\eta_c} \,  \eta_c(x) \, J_{\eta_c}(x),
\end{equation}
where $g_{\eta_c}$ is the coupling constant.
The current $J_{\eta_c}(x)$ is defined as
\begin{equation}
J_{\eta_c}(x) = \int d^4x_1 \int d^4x_2 \,  \delta^{(4)}\left( x-\dfrac{x_1+x_2}{2}\right)
\Phi_{\eta_c}\left[ (x_1-x_2)^2 \right]\bar{c}_a(x_1)\,  i\gamma^{5}\,  c_a(x_2).
\end{equation}
Similar to Eqs. \eqref{eq:phik} and \eqref{eq:phiXk}, $\phi_{\eta_c}$ denotes $\Phi_{\eta_c}$ in the momentum space,
\begin{equation}
\Phi_{\eta_c}(x^2)=\int \dfrac{ d^4k}{(2\pi)^4} \phi_{\eta_c}(k^2) \,  e^{-ik\cdot x}.
\end{equation}
$\phi_{\eta_c}$ is also chosen a Gaussian form,
\begin{equation}
\phi_{\eta_c}(k^2)=\exp(k^2/\Lambda_{\eta_c}^2).
\end{equation}
The parameter $\Lambda_{\eta_c}$ is chosen the fitted value $3.0$ GeV~\cite{dubnicka2010quark,PhysRevD.81.034010}.

$g_{\eta_c}$ can be obtained by the compositeness condition
\begin{equation}
1 - g_{\eta_c}^2 \,  \dfrac{d \Pi_{\eta_c}(s)}{ds}\bigg|_{s=m_{\eta_c}^2} = 0,
\end{equation}
where the function $\Pi_{\eta_c}(p^2) $ is defined as
\begin{equation}
\Pi_{\eta_c}(p^2)=i \int d^4x \,  e^{ip \cdot x}
\langle 0 | TJ_{\eta_c}(x)J_{\eta_c}(0)^{\dagger} | 0 \rangle.
\end{equation}

The analytical results of $g_{\eta_c}$ is
\begin{equation}
\begin{aligned}
	\dfrac{1}{g_{\eta_c}^2}=&
	\frac{3}{8 \pi^{2}}
\int_{0}^{\infty} d u_{1} \int_{0}^{\infty} d u_{2}
\ \frac{1}{ D_1^{5}}\
\left(
D_1^{2} D_2 \,  \widetilde{m}_1^{2} + 2D_1 \left( D_2 +  D_3 \right) + D_2 D_3 \,  \widetilde{m}_2^{2}
\right)
\exp\Big[\frac{D_2 }{D_1} \frac{\widetilde{m}_2^{2}}{2} - \widetilde{m}_1^{2}(u_1 + u_2)\Big],
\end{aligned}
\end{equation}
where the dimensionless mass parameters
\begin{equation}
\widetilde{m}_1=\dfrac{m_{c}}{\Lambda_{\eta_c}},  \quad
\widetilde{m}_2=\dfrac{m_{\eta_c}}{\Lambda_{\eta_c}},
\quad \widetilde{M}=\dfrac{m_{X}}{\Lambda_{\eta_c}}.
\end{equation}

\subsection{The decay width of $X(6900) \to 2\eta_c$}
The scattering amplitude of the decay $X(p) \to \eta_c(q_1) + \eta_c(q_2)$ can be also calculated with the same approach as $X(6900) \to 2J/\psi $. Because the above $X(6900) \to 2J/\psi $ result suggests the A-A coupling, the following amplitude $\mathcal{M}$ only adopts this coupling,
\begin{equation}
\mathcal{M} = \mathcal{M}_a+\mathcal{M}_b+\mathcal{M}_c
=
\varepsilon_{\mu\nu}^X(p)q_1^{\mu}q_1^{\nu}\dfrac{I}{m_X},\label{eq:dMetac}
\end{equation}
where $\mathcal{M}_a$, $\mathcal{M}_b$ and $\mathcal{M}_c$ come from the similar Figures~\ref{fig:X-decay1}--\ref{fig:X-decay3} ($J/\psi\to\eta_c$), respectively.
\begin{align}
\mathcal{M}_a ={}&
\frac{3}{32 \pi^{8}} \,  g_X g_{\eta_c}^{2} \,
\varepsilon_{\mu\nu}^X(p)
\int d^4 p_1 \,  d^4 p_2\,
\phi_{\eta_c}\left[ \Big(\dfrac{q_{1}}{2}+p_{1}\Big)^{2}\right]
\phi_{\eta_c}\left[ \Big(\dfrac{q_{2}}{2}+p_{2}\Big)^{2}\right]
\\
& \times
\phi_X\left(
\dfrac{8 p_1^2 + 8 p_1 \cdot q_1 + 8 p_2^2 + 8 p_2 \cdot q_2 + 3 q_1^2 - 2 q_1 \cdot q_2 + 3 q_2^2}{16}
\right)
\\
& \times
\mathrm{tr}\Big[S(p_1)  \gamma^5 S(p_1 + q_1) \gamma^\mu S(p_2) \gamma^5 S(p_2 + q_2) \gamma^\nu\Big],\label{eq:Metac}\\
\mathcal{M}_b={}&\mathcal{M}_c=0.
\end{align}
The scalar integrals $I$ can be found in Appendix \ref{subsec:SI}.

\begin{figure}[htbp]
\begin{minipage}[h]{0.4\textwidth}
\centering
\includegraphics[width=\textwidth]{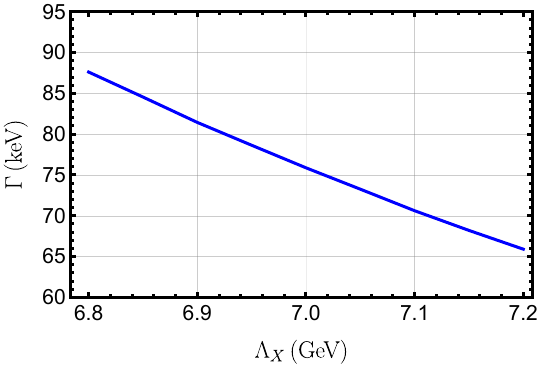}
\caption{Decay width of $X(6900)\to 2\eta_c$}\label{fig:etac}
\end{minipage}
\quad
\begin{minipage}[h]{0.5\textwidth}
\centering
\includegraphics[width=0.8\textwidth]{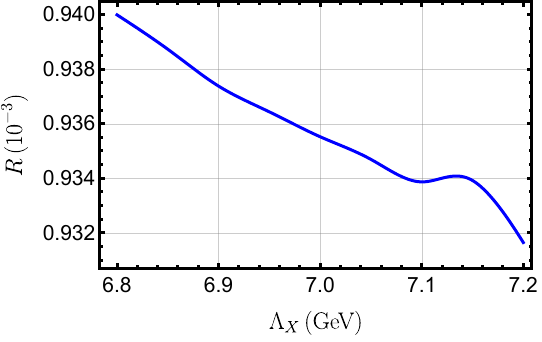}
\caption{Branching fraction $\dfrac{\Gamma(X(6900)\to 2\eta_c)}{\Gamma(X(6900)\to 2J/\psi)}$}
\label{fig:BR}
\end{minipage}
\end{figure}

The analytical decay width is
\begin{equation}
\Gamma=\dfrac{I^2}{120\pi}
\frac{(m_X^2/4 - m_{\eta_c}^2)^{5/2}}{m_X^4}.\label{eq:dwetac}
\end{equation}
Using a new input $m_{\eta_c}=2.9841$ GeV, the numerical value of $g_{\eta_c}$ is obtained to be 3.69 GeV, and the decay width shows in Fig. \ref{fig:etac}. It indicates the decay width of $X(6900)\to 2\eta_c$ is smaller than 100 keV, which is about one thousandth of the width of $X(6900) \to 2J/\psi$. It may be difficult to detect by experiment. The predicted branching fraction $\Gamma(X(6900)\to 2\eta_c)/\Gamma(X(6900)\to 2J/\psi)$ is presented in \cref{fig:BR}. Its value remains nearly constant at approximately $0.93 \times 10^{-3}$, although both $\Gamma(X(6900)\to 2\eta_c)$ and $\Gamma(X(6900)\to 2J/\psi)$ have a little larger changes.

Table~\ref{tab:com} presents a comparison between our results and others. Our $\Gamma(X(6900)\to 2\eta_c)/\Gamma(X(6900)\to 2J/\psi)$ is much smaller than others. To understand the significant difference between the decay widths of $\Gamma(X \to 2\eta_c)$ and $\Gamma(X \to 2J/\psi)$, a simple estimation based on Eqs. \eqref{eq:Gammaav} and \eqref{eq:dwetac} can be performed.
These two equations are reformulated in a more compact form
\begin{align}
\Gamma(X \to 2J/\psi)&=K_1(m_X^2/4 - m_{J/\psi}^2)^{1/2},\\
\Gamma(X \to 2\eta_c)&=K_2\frac{(m_X^2/4 - m_{\eta_c}^2)^{5/2}}{m_X^4 }.\label{eq:g2etac}
\end{align}
They lead to
\begin{align}
\frac{\Gamma(X \to 2\eta_c)}{\Gamma(X \to 2J/\psi)}=
\frac{K_2}{K_1}\frac{(m_X^2/4 - m_{\eta_c}^2)^{1/2}}{(m_X^2/4 - m_{J/\psi}^2)^{1/2}}\frac{(m_X^2/4 - m_{\eta_c}^2)^{2}}{m_X^4}.\label{eq:decayw}
\end{align}
$(m_X^2/4 - m_{\eta_c}^2)/(m_X^2/4 - m_{J/\psi}^2)\approx 1.3$, and the numerical result indicates that $K_1$ and $K_2$ are at the same order of magnitude. Hence, the greatest difference comes from the ratio $(m_X^2/4 - m_{\eta_c}^2)^{2}/m_X^4\approx 4\times 10^{-3}$. Performing a simple dimensional analysis, both Eqs. \eqref{eq:Mav-av} and \eqref{eq:Metac} have the same dimension. However, integrating Eq. \eqref{eq:Metac} leads to two external momenta $q_1^{\mu}q_1^{\nu}$, but Eq. \eqref{eq:Mav-av} does not have this parameter. To give a correct dimension, the leading term ($N_1$ term) in Eq. \eqref{eq:Mav-av} needs to multiply by $m_X$, but Eq. \eqref{eq:Metac} needs to divide by $m_X$ (see also Eq. \eqref{eq:dMetac}). There exists a parameter $m_X^2$ difference between these two decay amplitudes, i.e. a parameter $m_X^4$ difference for the decay width. The numerator $(m_X^2/4 - m_{\eta_c}^2)^{5/2}$ in Eq. \eqref{eq:g2etac} gives the correct dimension. It leads to the numerator $(m_X^2/4 - m_{\eta_c}^2)^{2}$ in Eq. \eqref{eq:decayw}. Therefore, the decay width of $X(6900)\to2\eta_c$ is significantly suppressed compared to that of $X(6900)\to2J/\psi$.

\section{SUMMARY AND DISCUSSION}
\label{sec:5}

This work employs the covariant quark model to calculate the decay width of $X(6900)\to 2J/\psi$, in order to investigate the internal structure of $X(6900)$.
$X(6900)$ is taken to be a tetraquark state with the diquark-antidiquark
structure $[cc][\bar{c}\bar{c}]$. Both A-A and V-V couplings are considered.
The model indicates that the A-A coupling is more consistent with the experiments.
The internal structure of $X(6900)$ is more likely an A-A $[cc][\bar{c}\bar{c}]$ coupling.
At least, if there is a mixing, the A-A coupling accounts for the majority of the contribution.
This work provides an additional support for the diquark-antidiquark configuration
as a viable description of exotic hadrons beyond the conventional quark model.
It also verifies the validity of the covariant quark model.
This model can reasonably extend to other exotic hadron states.

Furthermore, the decay width of $X(6900)\to 2\eta_c$ is also predicted in the range of $66\sim88$ keV.
Notably, the branching fraction between $X(6900)\to2\eta_c$ and $X(6900)\to2J/\psi$
decay channels is predicted about $0.93 \times 10^{-3}$.
It indicates that the $2J/\psi$ channel remains the dominant decay channel. Nevertheless, at present, there exists no experiment on $X(6900)\to2\eta_c$, this result needs to be verified in future experiments.

This paper only restricts $X(6900)$ to the compact diquark-antidiquark $[cc][\bar{c}\bar{c}]$
configuration, considering both V-V and A-A coupling schemes. As stated in the Introduction, $X(6900)$ is not possible to be a meson molecule. However, it may have other structures, such as hybrid state or mixing between different Fock components.

A possible hybrid form would be $(cc\bar{c}\bar{c}g)$, where the gluon provides an additional color-octet excitation~\cite{Wan:2020fsk}.
The QCD sum rule has studied this gluonic configuration on $J^{PC}=0^{++}$ and
$0^{-+}$, but a complete decay calculation for $J^{PC}=2^{++}$ has not yet been performed.

As an example, a Fock state mixing scheme would be $|X(6900)\rangle = \alpha|\text{tetraquark}\rangle +
\beta|\text{hybrid}\rangle$ (satisfying $|\alpha|^2 + |\beta|^2 = 1$). The decay width arises from these two components as well as the potential interference terms. Significant hybrid admixture would introduce some new decay pathways mediated by the additional gluon, depending on the mixing coefficient $\beta$. This typically leads to a variation in the predicted $2J/\psi$ partial width.
Although the study involving coupled-channel mixing or gluonic operators is beyond
the scope of this work, the aforementioned estimates suggest that as long as the compact
$[cc][\bar{c}\bar{c}]$ structure (particularly the A-A configuration) remains the dominant
contribution, even a modest hybrid admixture may only represent a minor correction to the current
result. To precisely quantify the mixing ratio $\beta$, we suggest that future research should
make a more detailed theoretical investigation of gluonic degrees of freedom.

\section{Acknowledgments}
This project was supported by the Guangxi Science Foundation under Grants No. 2025GXNSFAA069930, the Guangxi Science and Technology Innovation Platform Program (Leitai Action Plan, Grant No. Guike LT2600640026), Guangxi Key R\&D Program (Guangxi Funeng Action Plan, Grant No. Guike FN2504240040), and the ``Guangxi Highland of Innovation Talents'' Program.


\appendix
\section{The Scalar Integrals $R_1,R_2,N_1,N_2$}
\label{subsec:SI}

In the center-of-mass system, $q_1^2=q_2^2=m_{J/\psi(\eta_c)}^2,\, q_1 \cdot q_2 =m_{X}^2/2-m_{J/\psi(\eta_c)}^2$.

\begin{equation}
\begin{aligned}
N_1&=
\frac{3  }{4\pi^{4}M }
g_{X}\Lambda_X^3 g_{J/\psi}^{2}
\int_{0}^{\infty} d u_{1} \,
\int_{0}^{\infty} d u_{2} \,
\int_{0}^{\infty} d u_{3} \,
\int_{0}^{\infty} d u_{4}
\\
&\quad  \times
\dfrac{(4 C_{3}^{2} m_{1}^{2} + 8 C_{3} + C_{4} m_{2}^{2})(4 C_{2}^{2} m_{1}^{2} + 8 C_{2} + C_{5} m_{2}^{2}) }{C_{2}^{4} C_{3}^{4}}
\\
& \quad \times
\exp \Bigg[
\dfrac{m_2^2}{C_2C_3} \Big[(C_1+1) \lambda^4
+C_{6}\lambda^2
+ \dfrac{C_{7}}{4} \Big]
- C_{1} m_{1}^{2}
- \frac{M^{2}}{16}
\Bigg],
\end{aligned}
\end{equation}

\begin{equation}
\begin{aligned}
N_2=&
\frac{12}{\pi^4} \Lambda_X^3 g_X g_{J/\psi}^2 m_1^2M
\int_0^\infty\!\! d u_1
\int_0^\infty\!\! d u_2
\int_0^\infty\!\! d u_3
\int_0^\infty\!\! d u_4
\\
&\quad \times
\dfrac{ \exp
\left[
\dfrac{m_2^2}{C_2C_3} \Big[(C_1+1) \lambda^4
+C_{6}\lambda^2
+ \dfrac{C_{7}}{4} \Big]
- C_1 m_1^2 - \dfrac{M^2}{16}
\right] }
   { C_2^2 C_3^2 },
\end{aligned}
\end{equation}

\begin{equation}
\begin{aligned}
I=&
\frac{12}{\pi^4} \Lambda_X^3 g_X g_{\eta_c}^2 \widetilde{m}_1^2 \widetilde{M}
\int_0^\infty d u_1
\int_0^\infty d u_2
\int_0^\infty  d u_3
\int_0^\infty d u_4
\\
&\quad \times
\dfrac{ \exp
\left[
\dfrac{\widetilde{m}_2^2}{\widetilde{C}_{2}\widetilde{C}_{3}} \Big[(C_1+1) \widetilde{\lambda}^4
+C_{6}\widetilde{\lambda}^2
+ \dfrac{C_{7}}{4} \Big]
- C_1 \widetilde{m}_1^2 - \dfrac{\widetilde{M}^2}{16}
\right] }
   { \widetilde{C}_{2}^2 \widetilde{C}_{3}^2 },
\end{aligned}
\end{equation}
where
\begin{align}
\lambda &=\dfrac{\Lambda_X}{\Lambda_{J/\psi}}, \widetilde{\lambda}=\dfrac{\Lambda_X}{\Lambda_{\eta_c}},   \\[6pt]
C_{1}  &= u_{1} + u_{2} + u_{3} + u_{4},  \\[6pt]
C_{2}  &= 2\lambda^{2} + 2u_{1} + 2u_{3} + 1,\qquad \widetilde{C}_{2}= 2\widetilde{\lambda}^{2} + 2u_{1} + 2u_{3} + 1, \\[6pt]
C_{3}  &= 2\lambda^{2} + 2u_{2} + 2u_{4} + 1,\qquad \widetilde{C}_{3}= 2\widetilde{\lambda}^{2} + 2u_{2} + 2u_{4} + 1,  \\[6pt]
C_{4}  &= (2\lambda^{2} + 4u_{2} + 1)(2\lambda^{2} + 4u_{4} + 1),  \\[6pt]
C_{5}  &= (2\lambda^{2} + 4u_{1} + 1)(2\lambda^{2} + 4u_{3} + 1),  \\[6pt]
C_{6}  &= 2(u_{1}+u_{2}+u_{3}+u_{4})  + 2(u_{1}u_{2}+2u_{1}u_{3}+u_{1}u_{4}+u_{2}u_{3}+2u_{2}u_{4}+u_{3}u_{4}) + 1,  \\[6pt]
C_{7}  &= 3(u_{1}+u_{2}+u_{3}+u_{4})
+ 8(u_{1}u_{2}+u_{1}u_{3}+u_{1}u_{4}+u_{2}u_{3}+u_{2}u_{4}+u_{3}u_{4}) \\
     &\quad + 16(u_{1}u_{2}u_{3}+u_{1}u_{2}u_{4}+u_{1}u_{3}u_{4}+u_{2}u_{3}u_{4}) + 1.
\label{eq:CEF}
\end{align}

Formally, in the center-of-mass system,
\begin{equation}
R_1=2N_1,\quad
R_2=2N_2,
\end{equation}
but with different $g_X$.

\bibliography{ref}
\end{document}